\pdfoutput=1
\documentclass[aps,11pt,onecolumn,preprintnumbers,amsmath,amssymb,superscriptaddress]{revtex4}

\usepackage[english]{babel}
\usepackage{graphicx}
\usepackage{color}
\usepackage{epstopdf}

\begin{document}
\bibliographystyle{nature}
\newcommand{\unit}[1]{\:\mathrm{#1}}            
\newcommand{\To}{\mathrm{T_0}}

\newcommand{\Tp}{\mathrm{T_+}}
\newcommand{\Tm}{\mathrm{T_-}}
\newcommand{\EST}{E_{\mathrm{ST}}}
\newcommand{\Rp}{\mathrm{R_{+}}}
\newcommand{\Rm}{\mathrm{R_{-}}}
\newcommand{\Rpp}{\mathrm{R_{++}}}
\newcommand{\Rmm}{\mathrm{R_{--}}}
\newcommand{\ddensity}[2]{\rho_{#1\,#2,#1\,#2}} 
\newcommand{\ket}[1]{\left| #1 \right>} 
\newcommand{\bra}[1]{\left< #1 \right|} 
\renewcommand{\figurename}{\textbf{Figure}}

\title{Microsecond dark-exciton valley polarization memory in 2D heterostructures}

\author{Chongyun Jiang}
\affiliation{Division of Physics and Applied Physics, School of Physical and Mathematical Sciences, Nanyang Technological University, Singapore 637371, Singapore}
\author{Weigao Xu}
\affiliation{Division of Physics and Applied Physics, School of Physical and Mathematical Sciences, Nanyang Technological University, Singapore 637371,
	Singapore}

\author{Abdullah Rasmita}
\affiliation{Division of Physics and Applied Physics, School of Physical and Mathematical Sciences, Nanyang Technological University, Singapore 637371, Singapore}

\author{Zumeng Huang}
\affiliation{Division of Physics and Applied Physics, School of Physical and Mathematical Sciences, Nanyang Technological University, Singapore 637371, Singapore}

\author{Ke Li}
\affiliation{Division of Physics and Applied Physics, School of Physical and Mathematical Sciences, Nanyang Technological University, Singapore 637371, Singapore}
\author{Qihua Xiong}
\affiliation{Division of Physics and Applied Physics, School of Physical and Mathematical Sciences, Nanyang Technological University, Singapore 637371, Singapore}
\affiliation{NOVITAS, Nanoelectronics Center of Excellence, School of Electrical and Electronic Engineering, Nanyang Technological University, 639798 Singapore}
\affiliation{MajuLab, CNRS-Universit\'{e} de Nice-NUS-NTU International Joint Research Unit UMI 3654, Singapore}
\author{Wei-bo Gao}
\affiliation{Division of Physics and Applied Physics, School of Physical and Mathematical Sciences, Nanyang Technological University, Singapore 637371, Singapore}
\affiliation{MajuLab, CNRS-Universit\'{e} de Nice-NUS-NTU International Joint Research Unit UMI 3654, Singapore}
\affiliation{The Photonics Institute and Centre for Disruptive Photonic Technologies, Nanyang Technological University, 637371 Singapore, Singapore}


\begin{abstract}


\textbf{Transition metal dichalcogenides (TMDs) have valley degree of freedom \cite{XiaoLiuEtAl2012, Xu14}, which features optical selection rule  and spin-valley locking \cite{XiaoLiuEtAl2012, Xu14, MaK12,Zeng12,Cao12,Sallen12,Jones13}, making them promising for valleytronics devices and quantum computation \cite{LiEtAl2013, MakMcGillEtAl2014, GongLiuEtAl2013}. For either application, a long valley polarization lifetime is crucial.  Previous results showed that it is around picosecond in monolayer excitons \cite{Lagarde14, Cong14, Wang14},  nanosecond for electrons, holes or local excitons \cite{Yang15, Hsu15, Smolenski16, SrivastavaSidlerEtAl2015} and tens of nanosecond for interlayer excitons \cite{Xu16}. Here we show that dark excitons in 2D heterostructures provide a microsecond valley polarization memory thanks to the magnetic field induced suppression of valley mixing. The lifetime of the dark excitons shows magnetic field and temperature dependence which is consistent with the theoretical prediction.  The long dark exciton lifetime and valley polarization lifetime in 2D heterostructures make them promising for long-distance exciton transport and macroscopic quantum state generations.}
\end{abstract}

\maketitle

Reassembled layered Van der Waals heterostructures have revealed new phenomena beyond single material layers \cite{Geim13}. Especially, when two different monolayer TMDs are properly aligned, the electrons can be confined in one layer while holes are confined in the other layer \cite{Rivera15, Geim13, Xu16, XuWeigao2017}. Because the electron and hole wavefunctions in the two layers have little overlap, excitons can have a longer lifetime $\sim $ ns, as compared to $\sim$ ps for direct excitons. Such exciton is known as indirect exciton or interlayer exciton  \cite{Rivera15, Xu16}. A long exciton lifetime is crucial for high temperature macroscopically ordered exciton state, which forms the basis of a series of fundamental physics phenomena such as superfluidity \cite{Fogler14} and Bose-Einstein condensation \cite{Butov02-1, Butov02-2, High12}. Moreover, longer survival of excitons means longer distance of exciton transport, which is used for excitonic devices \cite{High08, Grosso09}.	 

In another perspective, similar to monolayer TMDs, the properly aligned 2D heterostructure has indirect exciton with valley degree of freedom. Valleys (K and K$^\prime$) are located at the band edges at the corners of the hexagonal Brillouin zone \cite{XiaoLiuEtAl2012, Xu14}.  The spins show opposite signs in the two valleys at the same energy, that is, spin-valley locking \cite{XiaoLiuEtAl2012, Xu14}. Moreover, these two valleys have opposite Berry curvature, leading to different optical selection rules in each valley. Through a circularly polarized optical pumping, they show the locking of the output photon chirality and valleys, introducing valley polarization \cite{MaK12,Zeng12,Cao12,Sallen12,Jones13}.  These unique properties put the valley degree of freedom as a possible candidate for opto-electronics and quantum computation with 2D material \cite{LiEtAl2013, MakMcGillEtAl2014, GongLiuEtAl2013}.  Leading to its application, a long valley polarization lifetime is a prerequisite and extensive efforts have been put towards particles and quasi-particles with longer lifetime in these ultrathin systems. Although previous lifetime measurement reveals that direct exciton lifetime is in the order of picosecond \cite{Lagarde14, Cong14, Wang14}, limiting its application to some extent, localized charge carriers can have a lifetime of around nanosecond \cite{Yang15, Hsu15, Smolenski16, SrivastavaSidlerEtAl2015} and indirect exciton can have tens of nanosecond valley polarization lifetime \cite{Xu16}.  On the other hand, experimental evidence shows the existence of dark excitons, lying tens of meV below the bright exciton in WSe$_2$ \cite{Xiaoxiao15} and their decay time is measured to be $\sim$ nanosecond in monolayer TMD \cite{Xiaoxiao16}. Here we report that, the dark excitons in 2D heterostructures can survive for microsecond timescale. With magnetic field suppressed valley mixing, they serve as a microsecond valley polarization memory for indirect excitons, which is 2 orders longer than the case without a magnetic field. In addition, we measured the $g$ factor $\sim 1$ for conduction band electrons  and dark excitons from magnetic field dependence dark exciton lifetime and dark exciton population in each valley.

%

\begin{figure}
\includegraphics[scale=0.4]{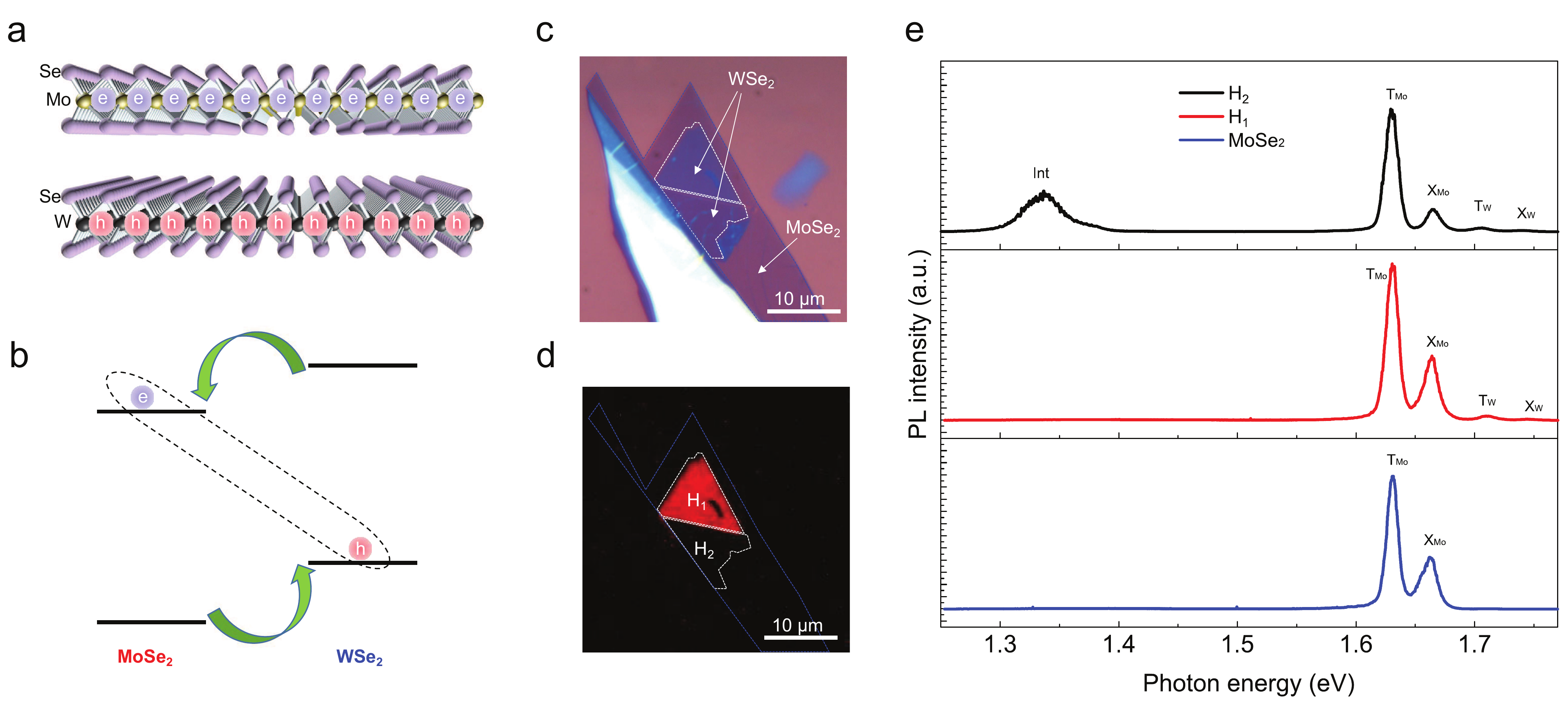}
\caption{ \textbf{Sample characterization.} \textbf{a}, MoSe$_2$ and WSe$_2$ form a 2D heterostructures, where electrons are confined in one layer and holes are confined in the other layer. \textbf{b}, Schematic of interlayer excitons. Electrons/holes in such 2D heterostructures tend to go to the ground states, which are the conduction band in MoSe$_2$ and valence band in WSe$_2$. Electrons and holes in different layer form the indirect excitons. \textbf{c}, Optical microscope image of the MoSe$_2$-WSe$_2$ heterostructure.  Blue dashed line shows the area of MoSe$_2$. White dashed lines show the region of heterostructure, which is separated into two area labeled as $H_1$ and $H_2$ in \textbf{d}. \textbf{d}, Fluorescence image taken with a
color camera, showing a bright ($H_1$) and dark ($H_2$) state in the two different place of heterostructure, under excitation with white light. \textbf{e}, Photoluminescence in the monolayer MoSe$_2$ and heterostructure $H_1$ and $H_2$ region under excitation with a laser with wavelength of 633 nm. Peaks on MoSe$_2$ are attributed to excitons ($X_{Mo}$) and Trions ($T_{Mo}$). In $H_1$ area, another two weak peaks appear and are labeled as excitons ($X_W$) and Trions ($T_W$) from WSe$_2$.  In $H_2$ area, another peak around $1.34 eV$ emerges, which is attributed to interlayer exciton and labeled as $Int$.    }\label{fig:1}
\end{figure}

Schematics and the optical spectroscopy of the MoSe$_2$-WSe$_2$ heterostructure on SiO2/Si substrate are shown in Figure \ref{fig:1}. In such heterostructuers, electrons tend to go to the conduction band of MoSe$_2$ and holes are confined in the valence band of WSe$_2$, forming the indirect excitons (Figure \ref{fig:1}a, \ref{fig:1}b). Our samples are prepared via a mechanical exfoliation and aligned-transfer method \cite{XuWeigao2017}. In this sample, the MoSe$_2$ monolayer is stacked on top of the WSe$_2$ monolayer. The detail of the sample preparation can be found in the Methods section. A fluorescence image of the heterostructure was taken with a color camera under white light excitation (Figure \ref{fig:1}d). It is found that the heterostructure consists of two areas: dark area with low intensity luminescence (labeled as $H_1$) and bright area with high intensity luminescence (labeled as $H_2$). The photoluminescence of these two areas as well as the MoSe$_2$ region under 633 nm laser excitation is shown in Figure \ref{fig:1}e. As can be seen from this figure, the interlayer exciton emission $\sim 1.34$ eV emerges for dark area $H_2$. The interlayer exciton emission intensity is comparable with the intralayer exciton and trion emission of MoSe$_2$. Interlayer emission is missing for the $H_1$ region with higher intensity of intralayer emission, which is due to the weak coupling between the two layers \cite{XuWeigao2017}. In the following measurement, we focused on  the interlayer exciton emission in $H_2$ area by using a 850 nm long pass filter in the photoluminescence (PL) collection.

\begin{figure}
\includegraphics[scale=1]{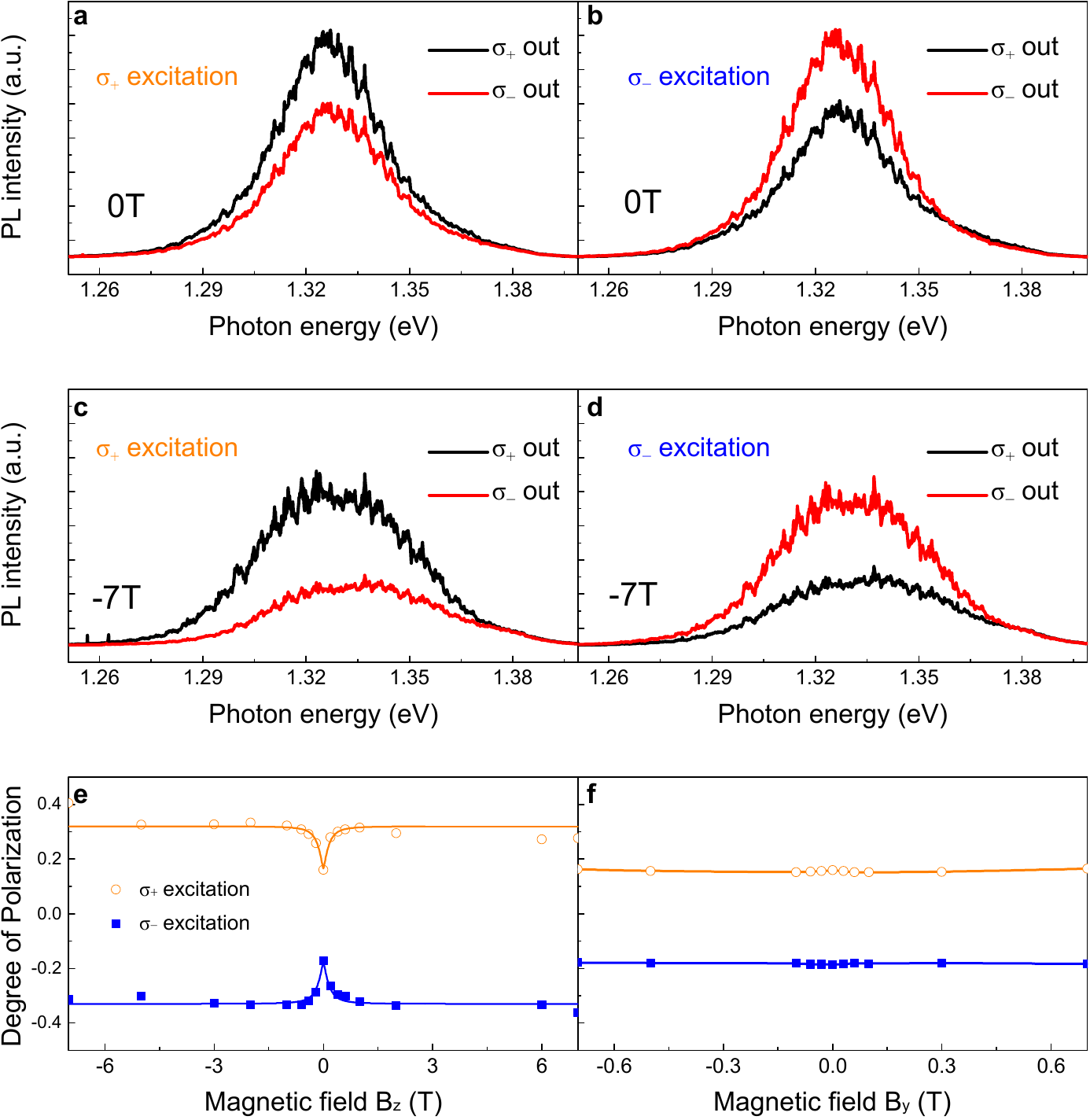}
\caption{ \textbf{Valley polarization with cw laser excitation. } \textbf{a, b}, Valley polarization at 0 T. Right and left circularly polarized light are labeled as $\sigma_+$ and $\sigma_-$. Under $\sigma_+$ laser excitation, $\sigma_+$ PL output component is more than $\sigma_-$ and vice versa for $\sigma_-$ excitation. This shows evidence of valley polarization.  \textbf{c, d}, Valley polarization at -7 T. Valley polarization is enhanced by applying magnetic field perpendicular to the sample surface. \textbf{e}, The valley polarization degree as a function of applied magnetic field in the z direction. The solid line is the fitting result following equation $P^j=P^j_0\pm P^j_1 (1-\frac{1}{r^2+r\sqrt{1+r^2}+1}), r= |B|/\alpha $ where $j$ indicates the excitation polarization, $P^j_0$ is the residual degree of polarization at 0 T due to valley polarization, $P^j_1$ is the saturation level of degree of polarization, and $\alpha$ represents the intervalley scattering between the dark exciton. 
\textbf{f}, The valley polarization degree as a function of applied magnetic field in the y direction with $B_z = 0$T. }\label{fig:2}
\end{figure}

We firstly carried out the measurement of valley polarization with continuous-wave (cw) pump laser, as shown in Figure \ref{fig:2}. 
In order to increase the count of interlayer exciton, the excitation laser has a wavelength of $1.708 eV$, corresponding to the resonant excitation of WSe$_2$ charged exciton. The polarization states of the excitation and detection are set to circularly polarization $\sigma_+$ or $\sigma_-$, and the degree of circular polarization is extracted from these four polarization combinations. The PL emission of different configurations at 0 T are shown in Figure \ref{fig:2}a-b. It is seen that the PL intensity of the co-polarization is always larger than that of the cross-polarization, corresponding to valley polarization. Next, we apply a magnetic field of -7 T perpendicular to the sample surface ($B_z$ direction). The results shown in Figure \ref{fig:2}c-d indicate that the difference between co-polarization and cross-polarization excitation gets larger at $B_z=-7$ T.

To quantify this, we performed the experiment of degree of polarization as a function of magnetic field in both $B_z$ (Faraday geometry) and $B_y$ directions (parallel to sample surface, Voigt geometry). Here we define the degree of polarization as $P^j=\frac{I_{\sigma_+}^j -I_{\sigma_-}^j}{I_{\sigma_+}^j+I_{\sigma_-}^j}$, where $I_{\sigma_+}^j$ ($I_{\sigma_-}^j$) is the $\sigma_+$($\sigma_-$) polarized PL intensity when excited with $j$ polarization. 
The individual degree of polarization pumped by $\sigma_+$ and $\sigma_-$ excitation versus magnetic field in Faraday and Voigt geometry are illustrated in Figure \ref{fig:2}e and \ref{fig:2}f. We can observe a dip of the degree of polarization at low magnetic field around 0 T in Faraday geometry. 
The valley polarization is around $17\%$ at 0 T magnetic field and quickly increases to $\sim 35\%$ at $\sim$1 T. For Voigt geometry, the degree of polarization does not show any dependence of the magnetic field. 

Similar with traditional semiconductor quantum well/dot, the valley depolarization in 2D material is caused by electron-hole exchange interaction \cite {Maialle93, bayer2002, Yu14}. The larger binding energy of excitons in monolayer TMDs further enhances such interaction, leading to valley depolarization and short valley polarization lifetimes. The intervalley scattering can be understood in term of in-plane depolarizing field \cite{Maialle93}. Hence, by increasing the magnitude of out-of-plane magnetic field the valley depolarization can be suppressed \cite{Smolenski16,bayer2002}. Following this, we fit the degree of polarization with equation $P^j=P^j_0\pm P^j_1 (1-\frac{1}{r^2+r\sqrt{1+r^2}+1}), r=  |B|/\alpha$ where $j$ indicates the excitation polarization, $P^j_0$ is the residual degree of polarization at 0 T, $P^j_1$ is the saturation level of degree of polarization, and $\alpha$ represents intervalley scattering term between the dark exciton. Experimental data fits very well with this model in small magnetic field range. The slight discrepancy at high magnetic field is attributed to valley relaxation through phonons, as we will discuss in the time resolved experiment.

\begin{figure}
\includegraphics[scale=0.7]{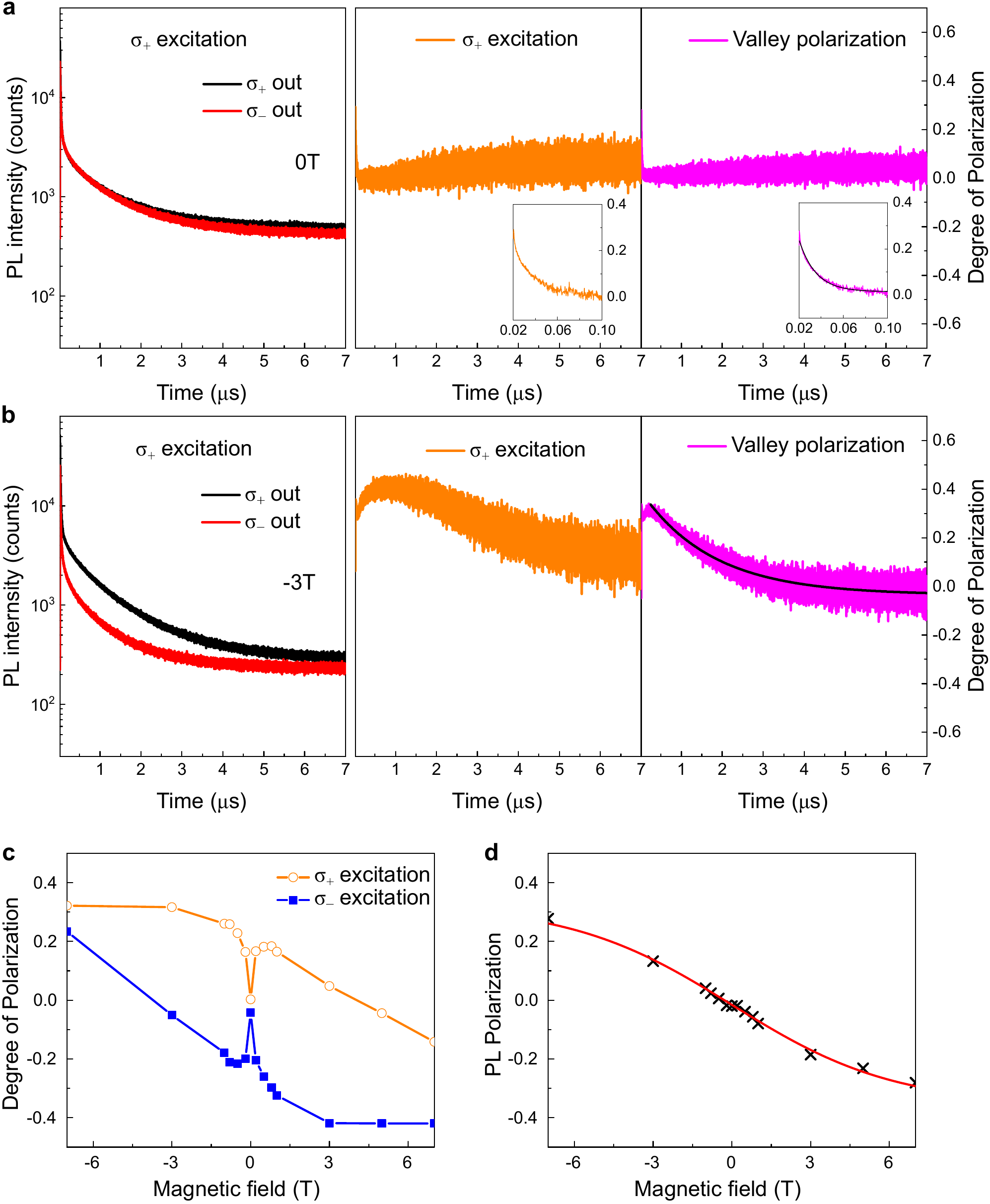}
\caption{ \textbf{Time resolved measurement for valley polarization and magnetic dependence of degree of polarization.} 
\textbf{a}, Time-resolved PL with $\sigma_+$ and $\sigma_-$ output under a pulsed laser excitation with polarization $\sigma_+$. The left panel shows the decay of the PL emission pumped by $\sigma_+$ excitation at 0 T while the middle and the right panel shows the calculated degree of polarization and valley polarization respectively.  The degree of polarization disappears quickly and hardly seen after $50$ ns and valley polarization has a decay time of $15\pm0.3$ ns. \textbf{b}, Similar as \textbf{a}, but at a magnetic field of -3 T in the z direction. The PL output difference between the two different polarization can be clearly seen even at $200$ ns. Valley polarization has a decay time of $1.745\pm0.007 \mu$s. \textbf{c}, Degree of polarization as a function of $B_z$ with pulsed laser excitation. The intensity is integrated from 21 ns to 7 $\mu$s. \textbf{d}, PL polarization as a function of $B_z$ with pulsed laser excitation. The PL polarization shows linear dependence on magnetic field for small magnetic field before saturating at big magnetic field.} \label{fig:3}
\end{figure}

To understand the dynamics of the interlayer exciton emission, we carried out time-resolved PL experiment with pulsed laser excitation. The laser has a repetition period 8 $\mu s$ and has the same wavelength as the cw experiment performed above. The left panels of the Figure \ref{fig:3}a and \ref{fig:3}b show the decay of the PL emission pumped by $\sigma_+$ excitation at 0 T and -3 T. The middle panels show the calculated  degree of polarization. We can see that, the degree of polarization $P^{\sigma_+}$ decays fast to zero at 0 T, while it has an extra slow decay component and remains above $0.2$ up to 2.5 $\mu s$ at -3 T. To qualify it, here we address the degree of polarization of two different types,  \textit{valley polarization} and \textit{PL polarization}. The former one depends on the polarization state of the excitation, i.e., copolarization and cross-polarization give different PL intensity. It can be calculated as $P_{val}=\frac{P^{\sigma_+}-P^{\sigma_-}}{2}$.
The latter one solely depends on the polarization of the PL emission, while not depending on the excitation, and it can be calculated as  the average of the individual degree of polarization pumped by $\sigma_+$ or $\sigma_-$ excitation  $P_{PL}=\frac{P^{\sigma_+}+P^{\sigma_-}}{2}$.
Right panels show the decay of valley polarization $P_{val}$ at 0 T and -3 T. valley polarization at 0 T has a decay time of $15\pm 0.3$ ns, while it has a decay time of $1.745\pm0.007 \mu$s at -3 T. More detailed PL data, valley polarization and PL polarization at -3 T, 0 T and 3 T are shown in Supplementary Figure \ref{timePL} and \ref{timedoP}.  

To study in detail the magnetic field dependence, we have measured the time-resolved PL when we vary out-of-plane magnetic field $B_z$. 
From the trace of PL, we integrate the intensity 7 $\mu s$ after the peak and calculate the degree of polarization $P^j=\frac{I_{\sigma_+}^j -I_{\sigma_-}^j}{I_{\sigma_+}^j+I_{\sigma_-}^j}$. As shown in Figure \ref{fig:3}c, the degree of polarization becomes larger when we apply a small magnetic field, similar with cw excitation; however, it becomes asymmetric when we change the magnetic direction in the upwards and downwards directions. It increases to a saturation level in one direction of magnetic field and in the other, it reduces after a certain magnetic field and got an inversed degree of polarization when $B_z$ is further enlarged. The tilting of curve is mainly caused by increased PL polarization $P_{PL}$ in the long timescale at high magnetic field, which is shown in Figure \ref{fig:3}d. The detailed fitting for PL polarization will be discussed after introducing the theoretical model.

\begin{figure}
 	\includegraphics[scale=0.6]{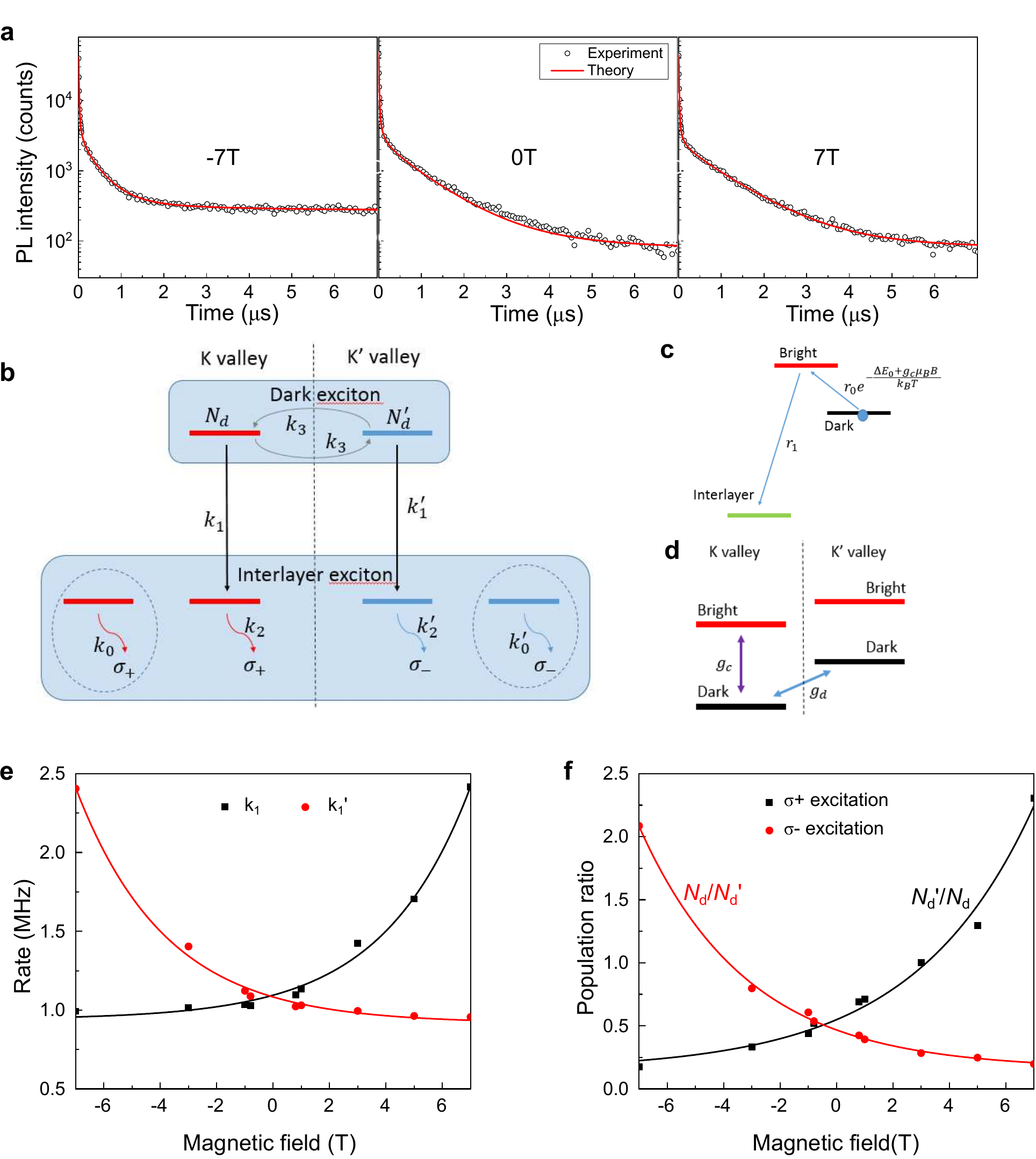}
 	\caption{\textbf{Theoretical model and experimental data fitting.} \textbf{a}, Sample of experimental data (temperature, $T = 2.3$K) and the fitting to the theoretical model. The data is obtained using $\sigma_-$ polarized pulsed excitation and $\sigma_-$ polarized PL detection. Log-linear plot is used. \textbf{b}, Theoretical model of the transition between exciton states in K and K' valley. Each valley has 1 dark exciton state and 2 interlayer exciton states. One of the exciton state population (circled in the figure) decays following power law with decay rate equal to $k_0$($k_0'$) times the corresponding state population. $N_d$ and $N_d'$ are the dark exciton population at K and K' valley respectively. $k_3$ is the intervalley scattering rate, $k_1(k_1')$ is the dark-to-interlayer scattering rate and $k_2(k_2')$ is the interlayer exciton decay rate. \textbf{c}, Model of magnetic field dependence of dark exciton-interlayer exciton conversion. Dark exciton can convert to interlayer exciton through intermediary bright exciton state. The magnetic field dependence of dark exciton-bright exciton energy difference (equal to $\Delta E_0$ at $B=0$ T) dictate the magnetic field dependence of dark exciton-interlayer exciton conversion. \textbf{d}, Dark and exciton energy level at finite magnetic field. The dark and bright exciton $g$ factor is denoted as $g_c$ while the one between dark excitons is $g_d$. \textbf{e}, Dark exciton-interlayer exciton transition rate ($k_1$) vs magnetic field. The dark exciton-interlayer exciton transition rates in both K valley ($\sigma$+ PL) and K' valley ($\sigma$- PL) are plotted against magnetic field and fitted using exponential function. \textbf{f}, The ratio between the dark exciton population at K and K'valley at different excitation polarization. The population ratio shows exponential dependence on magnetic field.}\label{fig:4}
 \end{figure}

Below we will analyse the origin of the long exciton lifetime and valley polarization lifetimes.  The results of the experimental data for $B = -7,0, 7$ T at low temperature ($T = 2.3$ K) in the case of $\sigma_-$ polarization excitation and $\sigma_-$ polarized PL detection are shown in Figure \ref{fig:4}a. First we consider the case at high magnetic field, where intervalley scattering is suppressed. The decay has both the slow and fast decay components. The slow decay in the order of $\tau_1 \sim 1 \mu$s suggests the dark exciton involvements, which will be further confirmed by the magnetic field dependence measurement shown below. The fast decay parts includes two parts, one of which is exponential decay and the other one is power-law decay. Hence, we fit the decay with three components as
\begin{equation}
I=A_1 e^{-\frac{t}{\tau_1}}+A_2 e^{-\frac{t}{\tau_2}}+\frac{B}{t+t_0},
\end{equation}
where $\tau_1$ and $\tau_2$ are related to the lifetimes in the slow and fast decay. $A_1$, $A_2$, $B$ and $t_0$ are other fitting parameters related to the initial population and rate constants. The value of $\tau_1$ is $\sim 1 \mu$s  while $\tau_2$ has a value of $\sim$ 10 ns. 
 
%


For the small magnetic field case, a more complete model that includes this intervalley scattering has to be used, instead of equation (1). A simplified version of this model showing the dark exciton of WSe$_2$ and the interlayer exciton energy levels is shown in Figure \ref{fig:4}b.  A more complete description are given in the supplementary information. The model is equivalent to equation (1) when the intervalley scattering rate, $k_3$ is negligible. As can be seen from Figure \ref{fig:4}a, both the experiment data for big and small magnetic field fits well with the theoretical model. One interesting finding is that the dark-to-interlayer exciton decay rate ($k_1$) is in the MHz regime. This means that the dark exciton lifetime is in the order of $\mu$s which is exceptionally long compared to the lifetime of other exciton types.


The $g$ factor of the conduction band electron can be obtained by studying the magnetic dependence of the dark exciton decay rate, $k_1$, at fixed temperature. In order to understand this, consider the process where the spin flipping happens before the interlayer charge transfer. In this case the dark exciton will transform to intermediate bright exciton before transforming to interlayer exciton (illustrated in Figure \ref{fig:4}c). According to previous measurement, charge transfer from bright monolayer exciton to interlayer exciton is very fast $\sim 100fs$ \cite{XuWeigao2017}. Therefore, we can safely neglect the charge transfer time, leading to the dark exciton relaxation rate $k_1=r_0e^{-(\Delta E_0+g_c\mu_BB)/k_BT}$, where $\Delta E_0$ is the energy difference between the dark exciton and bright exciton at 0 T. The other path of dark-to-interlayer exciton transition consist of the process where the interlayer charge transfer happens before the spin flipping. This process does not have strong magnetic dependence because the transition only happens from higher energy level to lower energy level. Hence, it will cause finite residual value of $k_1$ even at big magnetic field.

Dark exciton decay rate $k_1$ follows an exponential dependence on magnetic field  (Figure \ref{fig:4}e), which further confirms the correctness of this model and can be used to extract the $g$ factor for conduction band electrons.  For K valley, the value  $g_c = -1.11\pm0.095$ is found while it is equal to $1.07\pm0.079$ for K' valley. These $g$ factor values agree well with the theoretical prediction of the conduction band $g$ factor for WSe$_2$ when the out-of-plane effective spin $g$ factor has negative sign \cite{Kormanyos2014}. From the fitting in Figure \ref{fig:4}e, it can also be seen that the value of $k_1$ saturates to finite value $\sim1$ MHz at negative big magnetic field as predicted by the model. Additionally, the value of energy level difference between bright and dark exciton at zero magnetic field ($\Delta E_0$) can also be calculated from the temperature dependence of $k_1$. The detail of the experimental data and the theoretical fitting result related the temperature dependence of $k_1$ is shown in Supplementary Figure \ref{tempFit}.

Besides the conduction band electron $g$ factor, dark exciton $g$ factor for valley Zeeman splitting can be extracted from population of dark excitons in K and K' valleys (denoted as $g_d$ in Figure \ref{fig:4}d). Magnetic field will lift the energy degeneracy of dark excitons and cause the population difference between the two valleys.  As shown in Figure \ref{fig:4}f, the population ratio between the dark exciton in different valley fits well with exponential fitting which is the characteristic of Boltzmann distribution. The $g$ factor between the dark exciton in K and K' valley (denoted as $g_d$ in Figure \ref{fig:4}f) can be obtained from this fitting, which is found to be $0.9\pm0.04$. 

The lifting of the energy degeneracy of dark excitons is primarily the reason for observed PL polarization magnetic field dependence. At finite magnetic field,  the population ratio between these energy levels at thermal equilibrium will be determined by Boltzmann distribution. Because of this, PL polarization will have near-linear dependence at small magnetic field and saturate to $\pm1$ at big magnetic field. However, due to valley polarization, the magnitude of PL polarization will be less than 1. Following this line of reasoning, the PL polarization is fitted using equation $P_{PL} = P_{PL}^{s}\frac{1-e^{-\beta B}}{1+e^{-\beta B}}+P_{PL}^0$, where $\beta=\frac{g_v\mu_BB}{k_BT}$ with $g_v$ as the effective $g$ factor between K and K' valley, $P_{PL}^s$ is the saturation level of PL polarization, and $P_{PL}^0$ is a small residual PL polarization at 0 T due to experimental imperfection. From this fitting, we found the value of $g_v=1.1\pm0.09$. This is quite close to to the value of $g_d$, which indicates that dark exciton dynamics dominates the total PL polarization. 

In summary, we have experimentally demonstrated the long valley polarization lifetime in the order of microsecond in 2D heterostructures. This is primarily induced by magnetic field suppressed valley mixing for dark excitons. The long lifetime of the dark exciton serves as exciton reservoir for the interlayer exciton in long time scale. Both conduction band electron $g$ factor and dark exciton $g$ factor have been revealed from magnetic field dependence measurement. The survival of long exciton makes such 2D heterostructure promising for ultralong-distance exciton transport and exciton devices \cite{High08, Grosso09}. The possibilty to realize superfluidity \cite{Fogler14} in 2D heterostructures with long exciton lifetime may provide the future platform towards low-energy dissipation valleytronic devices.


\section*{Methods}

\subsubsection{Spectroscopy experiment setup}
 A homemade fiber-based confocal microscope is used for polarization-resolved PL spectroscopy. Polarizers and quarter wave plates are installed on the excitation and detection arm of the confocal microscope for polarization-selective excitation and PL detection. The PL emission is directed by an multi-mode optical fiber into a spectrometer (Andor Shamrock) with a CCD detector for spectroscopy recording. The sample is loaded into a magneto cryostat and cooled down to $\sim$ 2.3 K. Cryostat with vector magnet provides possibility to study dynamics in different magnetic field directions.  The vector magnetic field ranges from -7 T to +7 T in out-of-plane direction (z-axis) and -1 T to +1 T in in-plane direction (x- and y-axis). The wavelength of the excitation is 726 nm (1.708 eV) for both the cw and pulsed laser experiment (pulse width 100 ps).

\subsubsection{Preparation of the heterostructures}
We fabricated MoSe$_2$/WSe$_2$ heterostructures via a mechanical exfoliation and aligned-transfer method \cite{XuWeigao2017}. Bulk WSe$_2$ and MoSe$_2$ crystals (from HQ graphene) were used to produce WSe$_2$ and MoSe$_2$ monolayer flakes and they were precisely stacked with a solvent-free aligned-transfer process. We first prepared a WSe$_2$ monolayer on SiO2(300 nm)/Si substrate and a MoSe$_2$ monolayer on a transparent polydimethylsiloxane (PDMS) substrate; after careful alignment (for both relative position and stacking angle) under the optical microscope with the aid of an XYZ manipulation stage, we then stacked the two monolayer flakes together, forming a PDMS-MoSe$_2$-WSe$_2$-SiO2/Si structure. Finally, we removed the top PDMS layer and obtained a WSe$_2$/MoSe$_2$ heterostructure on SiO2/Si substrate. For controlled alignment of stacking angle, the armchair axes were guided according to their sharp edges from optical images, e.g., a stacking angle of $0(60)^\text o$ $(<\pm2^\text o)$ can be identified from Figure \ref{fig:1}a.

\section*{Supplementary Information}

\subsection{Analysis of time resolved degree of polarization}

The time resolved interlayer PL emission for $B=3,0,-3$T for $\sigma_+$ and $\sigma_-$ are shown in Supplementary Figure \ref{timePL}a and \ref{timePL}b respectively. Based on this measurement, the time resolved degree of polarization can be determined. The time-resolved degree of polarization, the valley polarization and the PL polarization are shown in Supplementary Figure \ref{timedoP}a, \ref{timedoP}b, and \ref{timedoP}c respectively. 

As can be seen from Supplementary Figure \ref{timedoP}, the lifetime of the valley polarization at 3T and -3T are similar to each other. In both cases, the valley polarization lifetime is much bigger than the 0T case. As has been explained in the main text, this is due to the suppression of the intervalley scattering when out-of-plane magnetic field is applied. This is not the case for PL polarization. The PL polarization at -3T differs from the one at 3T. In particular it saturates to different values at different magnetic field: $\sim0.2$ at -3T, and $\sim-0.2$ at 3T. This can be understood by the following scenario.


Initially, the polarized pulsed laser excites the bright exciton (trion) in one valley. This bright exciton population undergoes 4 main processes. It can:
\textbf{(1)} relax to vacuum which results in intralayer emission, \textbf{(2)} transform to interlayer exciton through charge transfer between MoSe2 and WSe2, \textbf{(3)} undergo intervalley scattering to the bright exciton in different valley, and \textbf{(4)} transform to dark exciton through spin flipping of the conduction band electron.

The first 3 processes happen in short time scale and they together determine the initial population of the interlayer exciton, which decays in tens of nanoseconds. This population decay is responsible for the fast decay of the valley polarization as shown in the inset of the main text Figure \ref{fig:3}a right panel. In the later stage ($> 50ns$), the dark exciton contribution to the interlayer exciton emission is the dominant process. In other words, the valley polarization and PL polarization of the interlayer exciton closely follow the dark exciton population decay dynamics. At more than 6 $\mu s$, since the thermal equilibrium has been reached, the degree of polarization has a weak dependence on the excitation polarization and it is determined mostly by the dark exciton population in the two valleys which follows Boltzmann distribution. Hence, the valley polarization saturates to near zero value and the PL polarization saturates to value determined by the dark exciton distribution between the two valleys. 

%

\subsection{Detail of theoretical model and experimental data fitting}

The diagram illustrating the theoretical model showing all the parameters involved in the fitting are shown in Supplementary Figure \ref{theoModel}. The interlayer exciton emission is proportional to the sum of the two interlayer exciton population. That is $I_{\sigma+} \propto (N_{xx}+N_x)$ and $I_{\sigma-} \propto (N_{xx}'+N_x')$.

The rate equation model can be solved analytically. The closed form solutions for $N_d, N_d', N_x, N_x', N_{xx},$ and $N_{xx}'$ are
\begin{equation}
N_d = c_1 e^{-tk_-}+c_2 e^{-tk_+}
\end{equation} 
\begin{equation}
N_x=c_3 e^{-tk_2}+c_1\frac{k_1}{k_2-k_-} e^{-tk_-}-c_2  \frac{k_1}{k_+-k_2} e^{-tk_+}
\end{equation} 
\begin{equation}
N_{xx}=\frac{1/k_0}{t+c_4 }
\end{equation}
\begin{equation}
N_d' = c_1 \frac{k_3}{\sqrt{\Delta^2+k_3^2}-\Delta}e^{-tk_-}-c_2 \frac{k_3}{\sqrt{\Delta^2+k_3^2}+\Delta}e^{-tk_+}
\end{equation} 
\begin{equation}
N_x'=c_5 e^{-tk_2'}+c_1\frac{k_3}{\sqrt{\Delta^2+k_3^2}-\Delta}\frac{k_1'}{k_2'-k_-} e^{-tk_-}+c_2\frac{k_3}{\sqrt{\Delta^2+k_3^2}+\Delta}  \frac{k_1'}{k_+-k_2'} e^{-tk_+}
\end{equation} 
\begin{equation}
N_{xx}'=\frac{1/k_0'}{t+c_6}.
\end{equation}
where $k_\pm=\frac{k_1+k_1'}{2}+k_3\pm\sqrt{\Delta^2+k_3^2}$ and $\Delta=\frac{k_1-k_1'}{2}$. The terms $N_d(N_d')$, and $N_x(N_x')$, are the population of $K(K')$ valley’s dark exciton and interlayer exciton.  $N_{xx}(N_{xx}')$ corresponds to the population of the additional exciton level at $K(K')$ valley to account for the power decay with decay rate equal to $k_0N_{xx}$ for $K$ valley and $k_0'N_{xx}'$ for $K'$ valley. $k_3$ is the intervalley scattering rate, $k_1(k_1')$ is the dark-to-interlayer scattering rate and $k_2(k_2')$ is the interlayer exciton decay rate.

It is possible to fit these equations straight away to the experimental data. However, due to the large number of the independent variables, there is a possibility that multiple local optimum solutions exist with some of the solution is not physically reliable. A different strategy can be employed by using the fact that for magnetic field with big enough magnitude the value of $k_3$ is negligible. In our case, we assume $k_3 = 0$ for $|B| = 0.8$, which is reasonable given the saturation of valley polarization at $|B| > 0.8$. In this case the interlayer population $N_x(N_x')$ and the dark exciton population $N_d(N_d')$ can be approximated as
\begin{equation}
N_d=c_2 e^{-tk_1}
\end{equation}
\begin{equation}
N_x=c_3 e^{-tk_2}+c_2\frac{k_1}{k_2-k_1} e^{-tk_1}
\end{equation}
\begin{equation}
N_d'=c_7 e^{-tk_1'}
\end{equation}
\begin{equation}
N_x'=c_5 e^{-tk_2'}+c_7\frac{k_1'}{k_2'-k_1'} e^{-tk_1'}.
\end{equation}

As can be seen from equation (7-10), for big magnetic field magnitude, there is no coupling between the exciton population in different valley. Hence, in this case, the model is equivalent to a time-dependent function with two exponentials and one power decay as stated in the main text. Moreover, due to the absence of the coupling terms, the experimental PL data for different emission polarization can be fitted separately. 

The fitting result at big magnetic field can be used to check the sanity of the model. The model predicts that the value of various decay rates ($k_1, k_1', k_2, k_2', k_0,$ and $k_0'$) and the population ratio between the interlayer exciton and dark exciton in one valley should be independent of the excitation polarization. We found that this is the case for all of these parameters other than the small difference for power decay rate ($k_0$ and $k_0'$). 

%

From the fitting result at big magnetic field, we extract the magnetic field dependence of $k_1, k_1', k_2, k_2', \frac{N_d}{N_x+N_{xx}}, \frac{N_d'}{N_x'+N_{xx}'}$, and  $\frac{N_d+N_x+N_{xx}}{N_d'+N_x'+N_{xx}'}$. From these data, the values of these parameters at small magnetic field are interpolated by linear interpolation. The value of $k_0, k_0', N_{xx}$, and $N_{xx}'$ are obtained by doing the fitting using equation (7-10) for both small and big magnetic field. The intervalley scattering rate ($k_3$) is used as a fitting parameter in equation (1-6) to fit the PL data for the case of small magnetic field. As shown in Figure \ref{fig:4}a in the main text, the fitting result fits well with the experimental data. 

%
%

\subsection{Temperature dependence of dark-to-interlayer exciton scattering rate}

In order to get the value of the energy level difference between bright and dark exciton at zero magnetic field ($\Delta E_0$), the pulse measurement were done in various temperatures at two different magnetic field setting: $B = 0$T and $B = 7$T. These data is then fitted with the theoretical model to obtain the value of dark-to-interlayer exciton scattering rate ($k_1$).

The measurement results for the case where $\sigma_-$ polarized pulsed excitation and $\sigma_-$ polarized PL detection can be seen in Supplementary Figure \ref{tempFit}a, \ref{tempFit}c. The corresponding temperature dependence of $k_1$ at two different magnetic field setting is shown in Supplementary Figure \ref{tempFit}b, \ref{tempFit}d. Based on this fitting, we obtain $\Delta E_0 = 58.2\pm20$ meV. Considering the experimental uncertainty, this result is comparable to the value of this parameter reported in \cite{Xiaoxiao15}.

\newpage

\bibliographystyle{nature}
\bibliography{hetero2DRef}

%
%
%

\newpage

\section{}

\renewcommand{\figurename}{\textbf{Supplementary Figure}}
\setcounter{figure}{0}

\begin{figure}
	\includegraphics[scale=0.9]{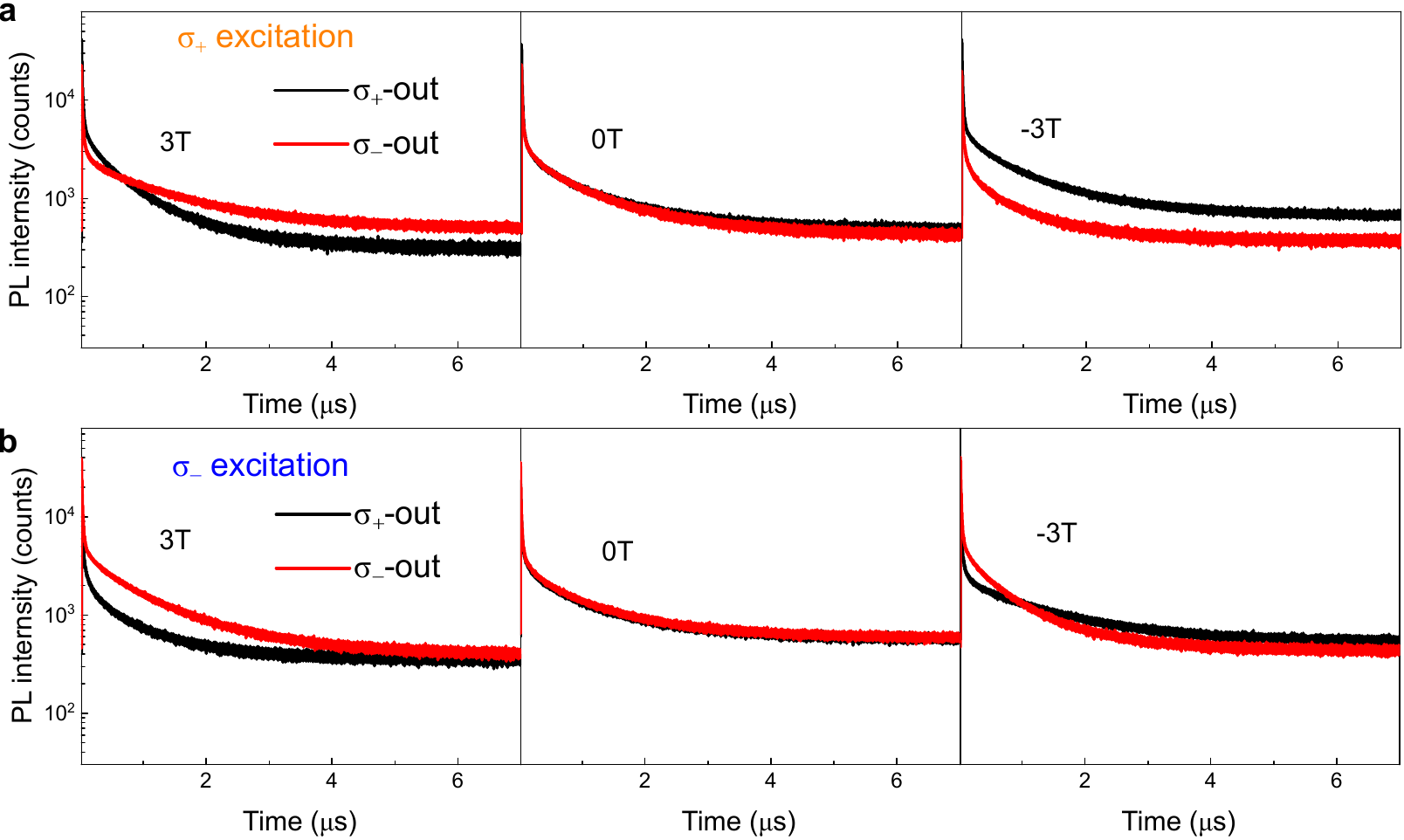}
	\caption{\textbf{Time resolved PL}. \textbf{a}, The three panels represent $\sigma_+$ excitation at out-of-plane magnetic field $B_z=$+3T, 0T and -3T.  \textbf{b}, $\sigma_-$ excitation at out-of-plane magnetic field $B_z=$+3T, 0T and -3T. }\label{timePL}
\end{figure}
\begin{figure}
	\includegraphics[scale=1]{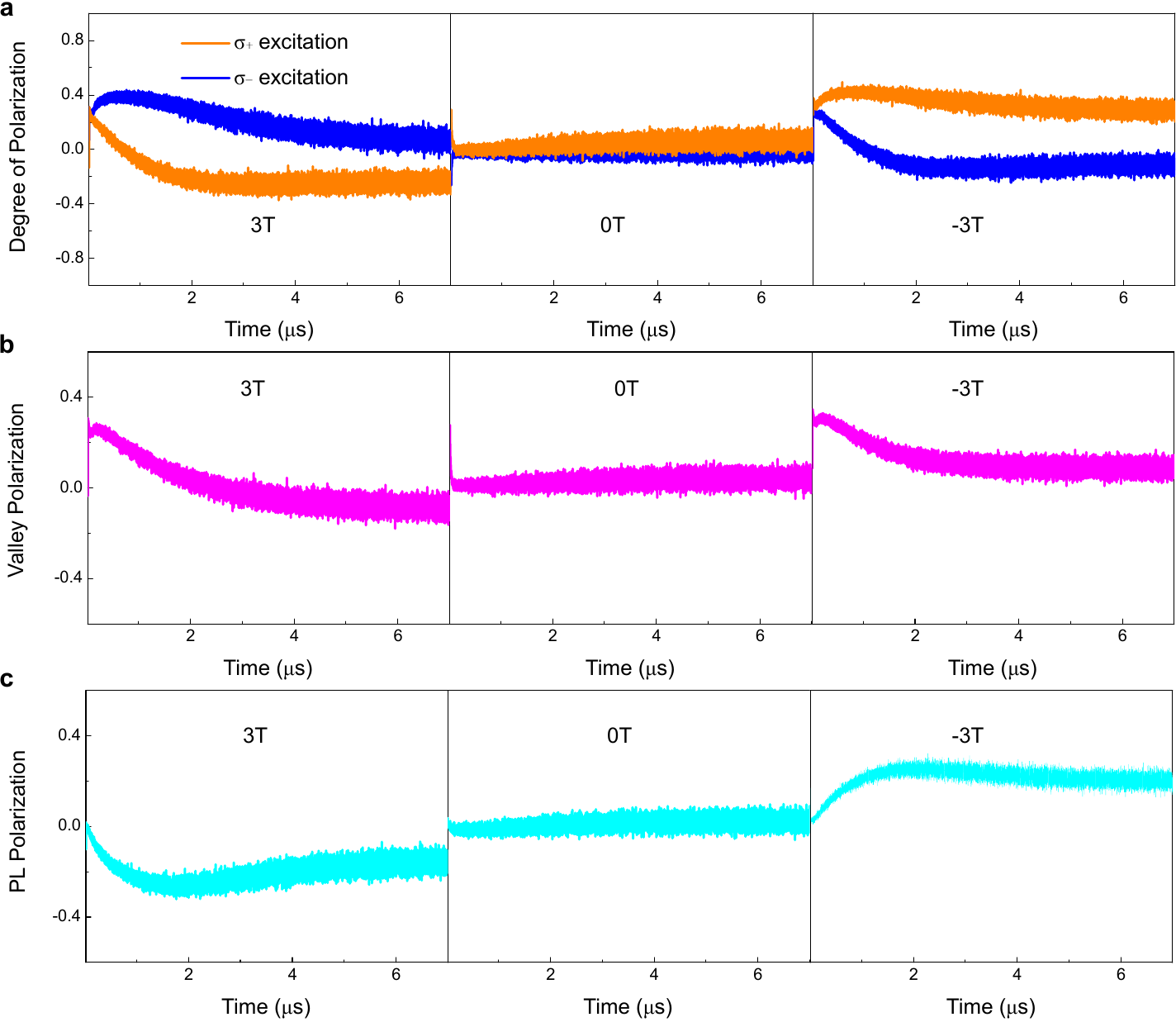}
	\caption{\textbf{Time resolved degree of polarization}. \textbf{a}, Degree of polarization $P^j$ for $\sigma_+$ excitation at out-of-plane magnetic field $B_z=$+3T, 0T and -3T. Here $P^j=\frac{I_{\sigma_+}^j -I_{\sigma_-}^j}{I_{\sigma_+}^j+I_{\sigma_-}^j}$,where $j$ indicates the excitation polarization and $I_{\sigma_+}^j$ ($I_{\sigma_-}^j$) is the $\sigma_+$($\sigma_-$) polarized PL intensity when excitation with $j$ polarization is used. \textbf{b}, Valley polarization $P_{val}=\frac{P^{\sigma_+}-P^{\sigma_-}}{2}$ at out-of-plane magnetic field $B_z=$+3T, 0T and -3T. \textbf{c}, PL polarization $P_{PL}=\frac{P^{\sigma_+}+P^{\sigma_-}}{2}$ at out-of-plane magnetic field $B_z=$+3T, 0T and -3T.}\label{timedoP}
\end{figure}

\begin{figure}
	\includegraphics[scale=0.5]{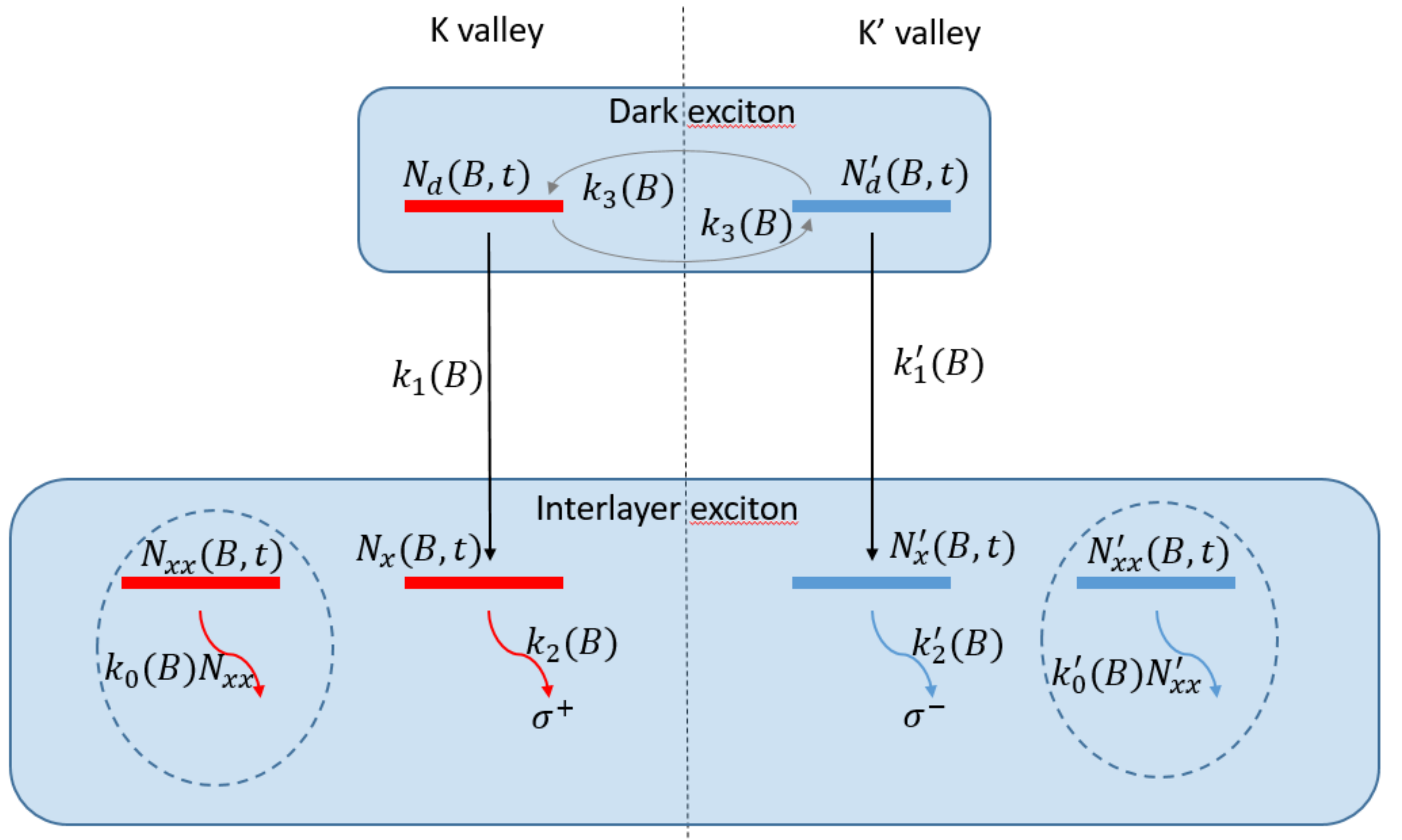}
	\caption{\textbf{Theoretical model and fitting parameters}. The term $N_d(N_d')$, and $N_x(N_x')$, are the population of $K(K')$ valley’s dark exciton and interlayer exciton.  $N_{xx}(N_{xx}')$ corresponds to the population of the additional exciton level at $K(K')$ valley to account for the power decay with decay rate equal to $k_0N_{xx}$ for $K$ valley and $k_0'N_{xx}'$ for $K'$ valley. $k_3$ is the intervalley scattering rate, $k_1(k_1')$ is the dark-to-interlayer scattering rate and $k_2(k_2')$ is the interlayer exciton decay rate. The term $B$ and $t$ denote the magnetic field and time dependence of the parameters.}\label{theoModel}
\end{figure}

\begin{figure}
	\includegraphics[scale=0.7]{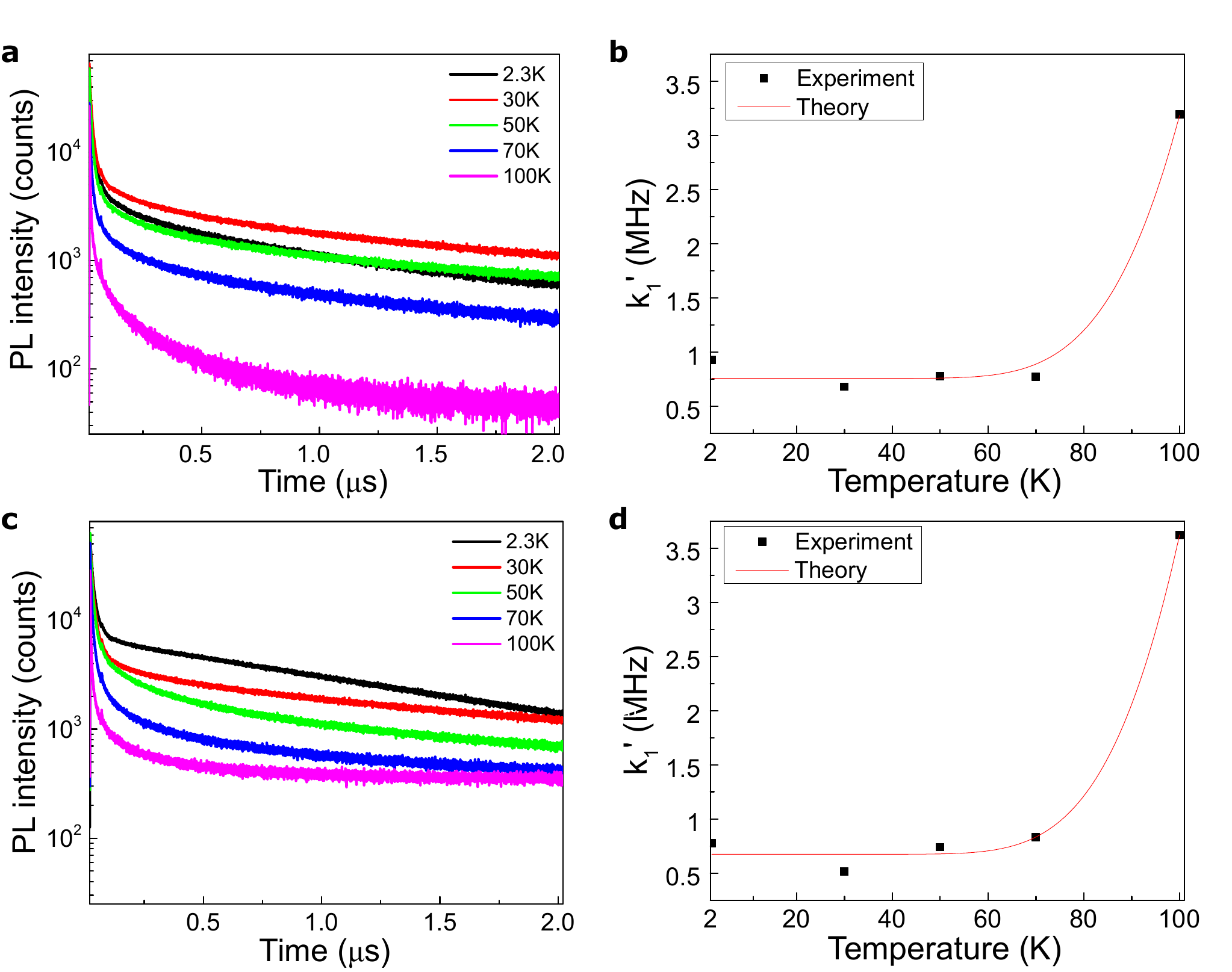}
	\caption{\textbf{Temperature dependence of the interlayer exciton PL}. The $\sigma_-$ polarized PL at $B = 0$T and  $B = 7$T are shown in (a) and (c) respectively. In both case, $\sigma_-$ polarized pulsed excitation were used. The temperature dependence of the corresponding dark-to-interlayer exciton scattering rate($k_1$) is shown in (b) and (d) respectively.}\label{tempFit}
\end{figure}

\end{document}